\documentclass[useAMS, usenatbib]{mn2e}
\usepackage{amsmath,color}
\usepackage{slantsc}
\usepackage{graphicx}

\newcommand{\ms}{ h^{-1}{\rm M_{\odot}}}
\newcommand{\mpch}{h^{-1}{\rm Mpc}}

\newcommand{\pch}{h^{-1}{\rm pc}}

\begin{document}
\bibliographystyle{mn2e}
\graphicspath{{./figs/}}
\title[]{Constraints on the identity of the dark matter from strong gravitational lenses}
\author[Ran Li et. al]
       {\parbox[t]{\textwidth}{
        Ran Li$^{1}$\thanks{E-mail:ranl@bao.ac.cn}, Carlos  S. Frenk$^{2}$, Shaun Cole$^{2}$,
        Liang Gao$^{1}$, Sownak Bose$^2$, Wojciech A. Hellwing$^{3,4}$
       }
        \vspace*{3pt} \\
  $^{1}$Key laboratory for Computational Astrophysics, Partner Group of the Max Planck Institute for Astrophysics, \\
  National Astronomical Observatories, Chinese Academy of Sciences, Beijing, 100012, China\\
  $^{2}$Institute for Computational Cosmology, Department of Physics, University of Durham, South Road, Durham, DH1 3LE\\
$^{3}$Institute of Cosmology and Gravitation, University of
 Portsmouth, Burnaby Road, Portsmouth PO1 3FX\\
  $^{4}$Interdisciplinary Centre for Mathematical and Computational
Modelling (ICM), University of Warsaw, ul. Pawi\'nskiego 5a, Warsaw, Poland\\
          }

\maketitle

\begin{abstract}
  The Cold Dark Matter (CDM) cosmological model unambigously predicts
  that a large number of haloes should survive as subhaloes when they
  are accreted into a larger halo. The CDM model would be ruled out if
  such substructures were shown not to exist. By contrast, if the
  dark matter consists of Warm Dark Matter particles (WDM), then below a threshold mass
  that depends on the particle mass far fewer substructures would be present.
  Finding subhaloes below a certain mass would
  then rule out warm particle masses below some value. Strong
  gravitational lensing provides a clean method to measure the subhalo
  mass function through distortions in the structure of Einstein rings
  and giant arcs. Using mock lensing observations constructed from
  high-resolution N-body simulations, we show that measurements of approximately {\color{black} 100 strong lens}
  systems with a detection limit of $M_{\rm low}=10^7\ms$ would clearly distinguish
  CDM from WDM in the case where this consists of 7~keV sterile neutrinos such as those
  that might be responsible for the 3.5~keV X-ray emission line
  recently detected in galaxies and clusters.
\end{abstract}

\section{Introduction}

A variety of observations indicate that dark matter accounts for more
than 80\% of the mass content of the Universe and so it dominates the
gravitational evolution of cosmic structure.  Its existence is
inferred through its gravitational effects in galaxies and clusters
and through the distortion of galaxy images by gravitational lensing
\citep[for a recent review see][]{Frenk2012}. Measurements of
temperature anisotropies in the Cosmic Microwave Background
\citep[CMB; e.g.][]{Planck2014} show that the dark matter is not
baryonic \citep[e.g.][]{Planck2014} but its identity remains unknown.

The CDM model in which the dark matter consists of cold collisionless
elementary particles (i.e. with negligible thermal velocities in the
early universe), such as the lightest stable supersymmetric particle,
has been shown, over the past 30 years, to provide an excellent match
to a variety of observations, many of them predicted in advance of the
measurements. These include the structure of the CMB temperature
anisotropies \citep{Peebles1982,Planck2014} and the pattern of galaxy
clustering \citep[][see \citealt{Frenk2012} for a comprehensive list
of references]{DEFW85, Springel2005, Tegmark2004,Cole2005}. There are
claims that the CDM particles may have already been detected through
$\gamma$-ray annihilation radiation from the Galactic Centre
\citep{HooperGoodenough2011} but these are controversial; the LHC has
not yet turned out any evidence for supersymmetry.

The Warm Dark Matter (WDM) model, in which the particles had
non-negligible thermal velocities at early times, is a viable
alternative to CDM. Indeed, there are also claims that such particles
may have been detected, in this case through particle decays resulting
in the 3.5~keV X-ray line recently discovered in galaxies and galaxies
clusters \citep{Boyarsky2014a,Bulbul2014a}.  A 7~keV sterile neutrino
originally introduced to explain neutrino flavour oscillations
\citep{Boyarsky2009} could be such a particle. However, these claims
are also controversial \citep[c.f.][]{Riemer2014}.

A very attractive feature of both the CDM and WDM models is that they
have predictive power; both are eminently falsifiable. The major
difference between them stems from the free-streaming cutoff in the
primordial power spectrum of density fluctuations which, in the case
of keV-mass particles, occurs on the mass scale of dwarf galaxies
whereas, in the case of cold particles, it occurs on the scale of
planets. Thus, on scales larger than individual bright galaxies, CDM
and WDM are almost indistinguishable, but on subgalactic scales they
make radically different predictions \citep[e.g.][]{Lovell2012, Kang2013,Bose2016, Ludlow2016}.

The most striking difference between CDM and WDM is the halo mass
function which turns over at the very different cutoff mass scales of
the two models. The halo mass function itself is difficult to measure
directly but, as we shall see in this paper, the mass function of
subhaloes (that is haloes that have been accreted into a larger halo and
survive) is accessible through observations. Rigorous and reliable
predictions for the halo and subhalo mass functions in CDM and WDM
exist from high-resolution N-body simulations
\citep{Springel2008,Gao2011,Colin2000,Avila-Reese2003,
  Lovell2012,Lovell2014,Hellwing2016,Cautun2014,Bose2016}. 
  
On the observational side, subhaloes can be detected through their gravitational
effects. Observations of the gaps in star streams can be used to find subhaloes 
within our own Galaxy \citep[e.g.][]{Erkal2015,Carlberg2012,Carlberg2013}; Gravitational lensing provides an powerful tool
to detect subhaloes outside the Milky way \citep[e.g.][]{Li2013,Mahdi2014, Li2014, Li2016,Hezaveh2014,Nierenberg2014}

Distinguishing keV-mass WDM from CDM requires
measuring the subhalo mass function (SHMF) below a mass of $\sim 10^{9}$ $\ms$. 
The most promising places to detect such subhaloes are the galatctic lenses.
The presence of subhaloes in the central regions of galactic haloes can perturb the
flux ratio of multi-image systems \citep[e.g.][]{Mao1998,
  Metcalf2001,Dalal2002}. It can also distort the images of extended giant arcs
or Einstein rings \citep[e.g.][]{Koopmans2005,Vegetti2009a,Vegetti2012,Hezaveh2016}.

Flux ratio anomalies have been measured only for a handful of quasars
and appear to reveal more small-scale structure than predicted even
for CDM, possibly due to projection effects from intervening haloes
and to inaccurate modelling of the complex mass distribution in the
lens galaxy \citep[e.g.][]{Xu2009,Xu2015}. For example, Hsueh et
al. (2015) have shown that the flux ratio anomaly of CLASS B1555+375,
one of the most anomalous lens systems known, can be explained by the
presence of a previously undetected edge-on disk in the lens
galaxy. 

Distortions of Einstein rings or giant arcs could offer a more direct
method. The technique developed by \citet{Koopmans2005} and
\citet{Vegetti2009a} can detect individual subhaloes and, using the
Bayesian formalism of \cite{Vegetti2009b}, a sample of detections can
constrain the SHMF.  \citet{Vegetti2014} analyzed 11 strong lenses in
the Sloan Lens ACS Survey \citep{Bolton2006} and obtained one
detection in SDSS J0956+5110. Their estimate of the projected
substructure mass fraction (i.e. the normalisation of the SHMF) is in
agreement with CDM and is lower than the values inferred from flux
ratio anomalies. However, the constraints on the slope of the SHMF
derived from such a small sample are weak. Many more strong lenses
will become available with future galaxy surveys such as Euclid and
LSST.

In this work we investigate how the detection of subhaloes in
perturbed Einstein rings or giant arcs can be used to distinguish the
SHMF in CDM and WDM. For this we make use of the high-resolution CDM
and WDM simulations of the Copernicus Complexio (\textsc{coco})
project \citep{Hellwing2016, Bose2016}. The \textsc{coco-warm}
simulation had an initial power spectrum appropriate to a thermal WDM
particle of 3.3~keV. It turns out that this power spectrum provides a
very good approximation to that of the coldest possible sterile
neutrino model that is compatible with the decay interpretation of the
3.5~keV X-ray line (corresponding to a value of the lepton asymmetry
parameter, $L_6=8.66$ \citealt{Lovell2015, Bose2016}).  Thus, ruling
out this particular model would exclude the entire family of 7~keV
sterile neutrinos.

The paper is organized as follows. In Section~\ref{sec:sim} we briefly
introduce the \textsc{coco} project.  In Section~\ref{sec:subdetect}
we estimate the probability of detecting subhaloes in dark matter halo
centres.   In Section~\ref{sec:bayesian} we present the modelling formalism of
subhalo detections.  In Section~\ref{sec:result} we show the constraining power of
subhalo detection from multiple lens systems on the SHMF. Our
conclusions are summarized in Section~\ref{sec:sum}

\section{Simulation data}
\label{sec:sim}
We use the \textsc{coco } simulations to derive the SHMF in a WDM
universe.  We begin by providing a brief discussion of the \textsc{coco }
simulations.

\subsection{$Copernicus$ $Complexio$ simulations}

The $COpernicus$ $COmplexio$ simulations \citep{Hellwing2016}, carried
out by the {\it Virgo Consortium}, consist of a set of cosmological
zoom-in simulations performed with a modified version of the Gadget-3
code \citep{Springel2001, Springel2005}. The region for resimulation
was extracted from the $Copernicus$ $Complexio$ $Low$ $Resolution$
(\textsc{color}) simulation (a periodic cubic volume of side
70.4$\mpch$); it contains 12.9 billion high resolution particles in a
roughly spherical region of radius 17.4$\mpch$. Each of the
high-resolution dark matter particles has a mass of $1.135\times
10^5\ms$. The gravitational softening was kept fixed at $230\pch$ in comoving unit.
Both \textsc{coco} and \textsc{color} assume the WMAP-7 cosmological
parameters \citep{Komatsu2011}: $\Omega_{\rm m}$ = 0.272,
$\Omega_{\Lambda}$= 0.728, $h$ = 0.704, $n_{\rm s}$ = 0.968 and
$\sigma_8$ = 0.81.

Simulations were performed for both a CDM and a 3.3~keV WDM universe:
\textsc{coco-cold} and \textsc{coco-warm} respectively.  The initial
conditions for both sets were arranged to have the same Fourier phases
and were generated using the method developed by \citet{Jenkins2013}.

The effect of free streaming at early times is to impose a cutoff in
the power spectrum. This is imposed in the initial conditions for \textsc{coco-warm}, 
through a modified transfer function, $T(k)$, so that the power
spectrum for WDM is related to that for CDM by:
\begin{equation}
P_{\rm WDM}(k)= T^2(k)P_{\rm CDM}(k) \,,
\end{equation}
where $T(k)$ is given by the fitting formula of \citep{Bode2001}:
\begin{equation}
T(k)=(1+(\alpha k)^{2\nu})^{-5/\nu} \,,
\end{equation}
where the constant, $\nu=1.12$,  and $\alpha$ depends on 
the WDM particle mass, $m_{\rm WDM}$, as 
\begin{equation}
\alpha= 0.049 \left ( \frac{m_{\rm WDM}}{\rm keV} \right)^{-1.11} \left ( \frac{\Omega_{\rm WDM}}{0.25}\right)^{0.11} \left( \frac{h}{0.7}\right)\mpch
\end{equation}
\citep{Viel2005}.  The smaller the WDM particle mass, the larger the
cutoff scale in the power spectrum cutoff.  In \textsc{coco-warm} the
equivalent thermal particle mass is $m_{\rm WDM}= 3.3$~keV. As
discussed in the introduction, this power spectrum is a very good
approximation to the power spectrum of the coldest possible sterile
neutrino model that is compatible with the decay interpretation of the
recently measured 3.5~keV X-ray line (corresponding to a value of the
lepton asymmetry parameter, $L_6=8.66$; \citealt{Lovell2015,
  Bose2016}). This power spectrum leads to a delay in the formation
epoch of haloes of mass below $\sim 2\times10^9\ms$ in \textsc{coco-warm}
relative to \textsc{coco-cold} \citep{Bose2016}.  We refer the reader
to \citet{Bose2016} and \citet{Hellwing2016} for further details of
the \textsc{coco} simulations.

\subsection{Subhaloes in COCO-WARM and COCO-COLD}

Haloes in the \textsc{coco }simulations were identified using the FOF
algorithm \citep{DEFW85} with a linking length of 0.2 times the mean
interparticle separation. Gravitationally-bound subhaloes within each
halo were identified using the SUBFIND algorithm
\citep{Springel2001}. Since the initial conditions for both
\textsc{coco-warm} and \textsc{coco-cold} had the same initial Fourier
phases, any differences in the abundance of low mass subhaloes between
the two are due entirely to the different input power spectra.

In order to obtain the true mass function in WDM simulations, it is
necessary to identify and exclude artificial haloes that form in
N-body simulations from initial power spectra with a resolved cutoff,
as is the case for \textsc{coco-warm}. These spurious, small-mass
haloes are generated by discreetness effects that cause fragmentation
of filaments, as discussed by \citet{Wang2007} in the context of
simulations from hot dark matter initial conditions. The same
phenomenon is seen in WDM simulations \citep{Angulo2013, Lovell2014,
  Bose2016}.  \citet{Wang2007} found that a large fraction of these
spurious haloes can be removed by eliminating haloes with mass below,
\begin{equation}
M_{\rm lim} = 10.1\, \bar{\rho}\, d\,k_{\rm peak}^{-2} \,,
\end{equation}
where $d$ is the mean interparticle separation and $k_{\rm peak}$ the
wavenumber at which the dimensionless power spectrum,
$\Delta(k)^2=\frac{k^3}{2\pi^2}P(k)$, reaches its maximum. Spurious haloes can also
be identified by tracing back their particles to the (unperturbed)
initial density field. The Lagrangian regions from which spurious
haloes form tend to be much flatter that the corresponding region for
genuine haloes \citep{Lovell2014}.  By calculating the inertia tensor
of the initial particle load, the sphericity of a halo can be defined
as $c/a$, where $a^2$ and $c^2$ are the largest and smallest
eigenvalues of the inertia tensor. Spurious haloes in the
\textsc{coco-warm} catalogues were removed by \citet{Bose2016} by
eliminating all haloes with $s_{\rm half-max}<0.165$ and $M_{\rm max}
< 0.5 M_{\rm lim}$, where $s_{\rm half-max}$ is the sphericity of the
halo at the half-maximum mass snapshot and $M_{\rm max}$ is the
maximum mass a halo achieved during its growth history. 

{\color{black} Note that, the halo selection in WDM is sensitive to these criteria.  
In \citet{Bose2016}, the sphericity cut is calibrated with respect to CDM simulations and the 
maximum mass cut is calibrated by matching simulations of different resolution.
We refer the reader to \citet{Lovell2014} and \citet{Bose2016} for a detailed
discussion.}

\begin{figure}
\includegraphics[width=0.5\textwidth]{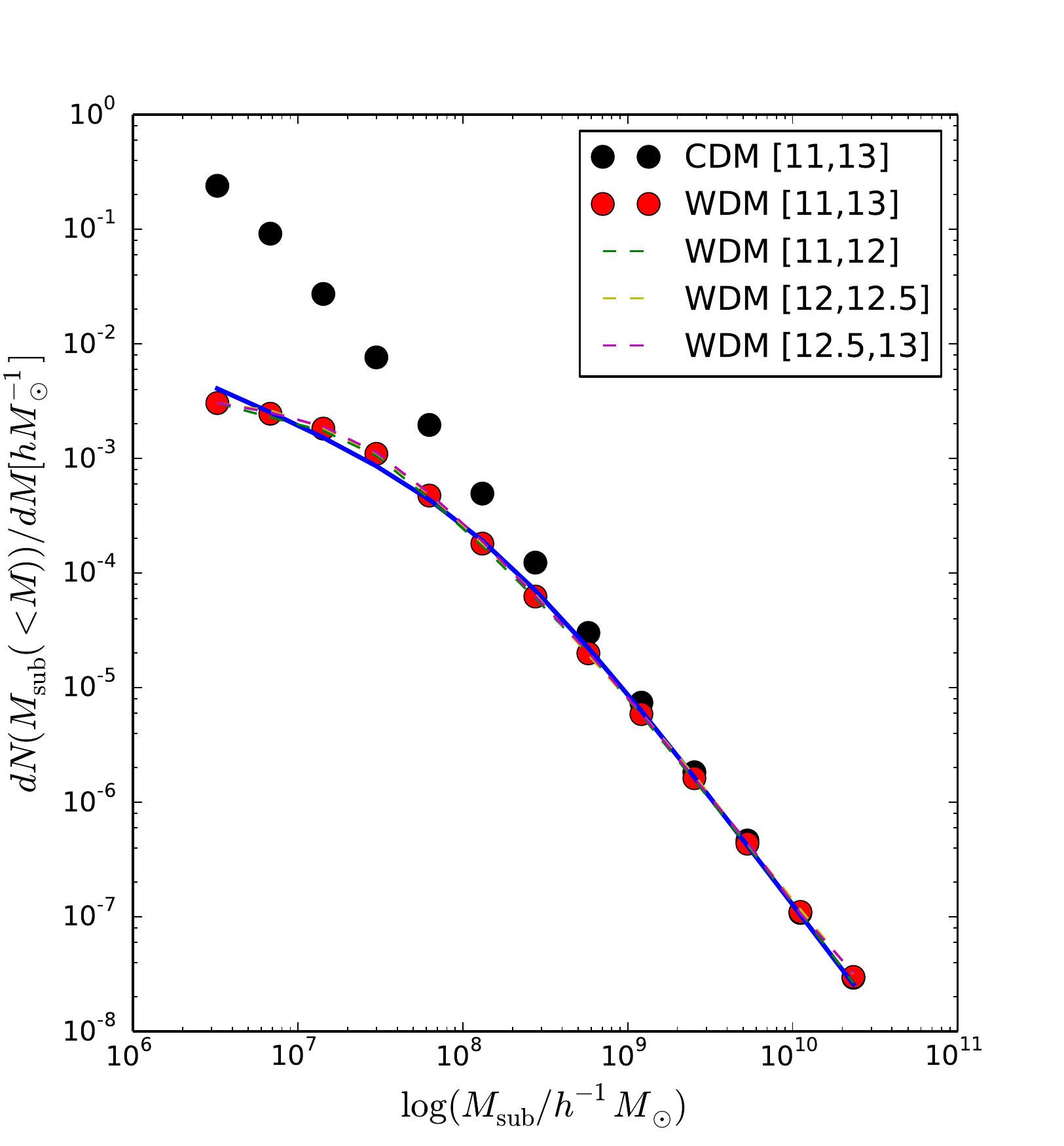}
\caption{The differential subhalo mass function for host haloes of
  different mass.  The solid black (\textsc{coco-cold}) and red
  (\textsc{coco-warm}) points show the subhalo mass function for host
  haloes with mass in range $[10^{11}\ms,10^{13}\ms$]. The dashed
  lines show the mass function for haloes in different mass ranges in
  \textsc{coco-warm}. We scale the dashed lines to match the red solid
  points (by requiring the average amplitude of different curves to be
  the same), so that one can compare the shape of the subhalo mass
  functions in \textsc{coco-warm}. The blue line shows the fit to the
  SHMF in \textsc{coco-warm}.}
\label{fig:SHMF}
\end{figure}

In Fig.~\ref{fig:SHMF}, we show the differential subhalo mass function
(SHMF) in COCO simulations. The SHMF
in \textsc{coco-cold} can be fitted by the power law, 
$n(M)\equiv dN(<M_{\rm sub})/dM=A_0 M^{-\alpha}$, where $N(<M_{\rm
  sub})$ is the total number of subhaloes with mass smaller than
$M_{\rm sub}$ and $\alpha=1.9$ \citep{springel2009,
 Gao2012}. The \textsc{coco-warm} simulation produces similar numbers of
subhaloes as \textsc{coco-cold} at larger masses but much smaller
numbers for   $M_{\rm sub} > 10^9\ms$. 
The slope of the SHMF in \textsc{coco-warm} begins to deviate
appreciably from $\alpha=1.9$ at $\sim 10^8\ms$. At at $10^7\ms$, 
the difference between the two SHMFs has grown to be a factor of 10. 
{\color{black} In Fig.~\ref{fig:SHMF}, we plot SHMFs in host haloes of different mass bins, and
find that they all have the same shape. The SHMF in \textsc{coco-warm} can be fitted with the
expression used by  \citet{Schneider2012}:}
\begin{equation}
\label{eq:nw_nc}
n_{\rm WDM}/n_{\rm CDM} = (1+m_c/m)^{-\beta} \,.
\end{equation}
\citet{Lovell2014} show that the WDM mass function is well fit
adopting $\beta=1.3$.  We fix $\beta=1.3$ and fit the mass function of
\textsc{coco-warm} to find a best-fit value of $m_c=1.3\times10^8
\ms$. The corresponding fit is shown by the solid lines in the
Fig.~\ref{fig:SHMF}.

\section{Substructure detection in strong gravitational lenses}
\label{sec:subdetect}

If the projected position of a subhalo is close to the Einstein radius
of a strong lens system, it can perturb the surface brightness
distribution of the Einstein ring. The strength of the perturbation
depends on the mass of the subhalo and its relative distance to the
Einstein ring.

To investigate the probability of a subhalo falling in the region of
an Einstein ring, we first calculate the Einstein radius of 
dark matter haloes of a given mass. In the real Universe, the size of
the Einstein radius is determined by the central mass distribution 
which, in sufficiently large haloes, is dominated by the
baryonic component of the galaxy. Previous analyses have shown that
modelling the total central mass distribution as a
singular-isothermal-sphere (SIS) can successfully predict the location
of strong lensing images \citep[e.g.][]{Koopmans2006, Gerhard2001,
  Czoske2008}.

Denoting the stellar velocity dispersion as $\sigma_v$, the Einstein
radius of a SIS can be written as:
\begin{equation}
\theta_E=\frac{4\pi\sigma_v^2}{c^2}\frac{D_{\rm l,s}}{D_{\rm s}},
\label{eq:RE}
\end{equation}
where $D_{\rm s}$ is the angular diameter distance from the source to
the observer and $D_{\rm l,s}$ is the distance between the lens and
the source. Since \textsc{coco} is a set of dark matter-only
simulations, it provides halo masses but not stellar velocity
dispersions. A convenient way to infer the latter is to take them from
the stellar velocity-dispersion {\em vs} halo-mass relation obtained
in a realistic cosmological hydrodynamics simulation. Here we use the
recent \textsc{eagle} reference simulation which follows the coupled
evolution of baryons and dark matter in a cubic volume of side
100~Mpc, with gas mass resolution of $1.8\times10^6 {\rm M_{\odot}}$
and softening length of 0.7 kpc \citep[][]{Schaye2015}. \textsc{eagle}
provides a good match to both the observed stellar mass function and
the galaxy size-stellar mass relation so it is reasonable to assume
that the stellar velocity dispersions are also realistic. Using the
public \textsc{eagle}
  database\footnote{http://www.eaglesim.org/database.html}
  \citep{McAlpine2015}, we find that the velocity-dispersion {\em vs}
  halo-mass relation is well fit by:
\begin{equation}
\sigma_v=\sigma_0\frac{(M/M_1)^{\gamma_1}}{(1+M/M_1)^{\gamma_2-\gamma_1}} \,,
\label{eq:vdisp}
\end{equation}
where $M$ is the halo mass, $\sigma_0=117$ km $\rm s^{-1}$,
$M_1=1.5\times10^{12} \ms$, $\gamma_1=4.30$, $\gamma_2=6.79$, and
$\sigma_v$ is the average stellar velocity dispersion within the inner
5~kpc of the central galaxy.  

\citet{Vegetti2014} have shown that the probability of detecting a
substructure in an Einstein ring depends on the mass and position of
the subhalo and on the gradient of the surface brightness distribution
of the lensed galaxy. In this work, we adopt the simple assumption
that within a thin region around the Einstein ring, any subhalo of
mass larger than a threshold, $M_{\rm low}$, can be detected through
its perturbation to the Einstein ring \citep{Vegetti2009b}.  In a
forthcoming paper we will investigate the effect of a more realistic
sensitivity function based on the results of \citet{Vegetti2014}.
Following \citet{Vegetti2009b}, we take the width of this thin annulus
to be 2$\Delta \theta=0.6$ arcsec.

The dark matter mass contained in the Einstein ring, $M_{\rm ring}$,
is given by:
\begin{equation}
\label{eq:MR}
M_{\rm ring}(R_E)=\int^{R_E+\Delta R }_{R_E-\Delta R } 2\pi R\,
\Sigma_{\rm dm}(R) \, dR \,,
\end{equation}
where the Einstein radius, $R_E=\theta_E D_{\rm l}$; $\Sigma_{\rm
  dm}(R)$ is the surface mass density of the dark matter halo; and
$\Delta R=\Delta\theta D_{\rm l}$.

From Eq.\ref{eq:nw_nc}, the probability of finding a subhalo of 
mass, $m$,  per unit volume can be written as:
\begin{equation}
\label{eq:Pm}
\frac{dP}{dm}\Big\vert_{\rm true}=A_0m^{\alpha} (1+m_c/m)^{-\beta} \,, 
\end{equation}
where for \textsc{coco-warm}, we have $\beta$=1.3 and 
  $m_c=1.3\times10^{8}\ms$, whereas for \textsc{coco-cold}, $m_c=0$.

  We denote the maximum and the minimum mass of the subhaloes of
  interest that lie within the Einstein ring region as $M_{\rm max}$
  and $M_{\rm min}$ respectively and adopt $M_{\rm max}=10^{10}\ms$ and
  $M_{\rm min}=10^6\ms$. We can then define a normalization factor, $A_0$, as:
\begin{equation}
A_0=\frac{1}{\int^{M_{\rm max}}_{M_{\rm min}} m^{\alpha}(1+m_c/m)^{-\beta} dm}.
\end{equation}
The expectation value of the number of  subhaloes in the Einstein ring region with mass $M_{\rm min}<m<M_{\rm max}$
can then be written as:
\begin{equation}
\mu_0(\alpha,\beta,m_{\rm c},f_E, M_{\rm ring})=\frac{f_E M_{\rm ring} }{\int^{M_{\rm max}}_{M_{\rm min}} m \frac{dP}{dm}\Big\vert_{\rm true} dm}\,,
\end{equation}
where $f_E=f_{\rm sub}(R_E)$ and $f_{\rm sub}$ is the fraction of mass contained in subhaloes at a projected radius $R$. 

When a halo merges into a larger system and becomes a subhalo, it
experiences dynamical friction and tidal striping. Subhaloes spiral
into the centre of the host halo and loose mass and many of them are
completely disrupted. As a result, we expect the fraction of mass
contained in subhaloes to increase with projected radius.
The \textsc{coco} volume contains only a few dark matter haloes of 
mass larger than $10^{13}\mpch$, making the estimation of $f_{\rm
  sub}$ noisy. We therefore make use of the analytical formula for
$f_{\rm sub}(R)$ derived by \citet{Han2015}. For dark matter haloes of
mass in the range $[10^{13}, 10^{14}]$ $\ms$, $f_{\rm sub}$ can be
approximated as:
\begin{equation}
f_{\rm sub}=0.35(R/r_{\rm vir})^{1.17}\,,
\label{eq:fr}
\end{equation}
 where $r_{\rm vir}$ is the virial radius of the halo and $R$ the projected radius.

 Observationally, it is only possible to detect subhaloes more massive
 than a certain threshold. \citet{Vegetti2009b} found the measurement
 errors on subhalo mass to be approximately Gaussian distributed with
 standard deviation, $\sigma_m$. In our catalogues, we will consider
 as `detected subhaloes' those having a measured mass larger than
 $M_{\rm low} \equiv 3\sigma_m$. We note that this definition is
 different from that adopted by \citet{Vegetti2014}, who empolyed a
 detection threshold derived from the probability density of a
 substructure mass, given the observed lensed data, marginalised over
 the host lens and background source parameters.

 Taking into account the detectability of a subhalo, we can rewrite
 the expected number of subhaloes in the Einstein ring region as:
\begin{eqnarray}
\begin{split}
&\mu(\alpha,\beta,m_{\rm c},f_E,M_{\rm ring} )= \\
 &\mu_0\int^{\infty}_{M_{\rm low}} \int^{M_{\rm max}}_{M_{\rm min}} \frac{dP}{dm}\Big\vert_{\rm true} \exp \left[\frac{(m-m')^2} {2\sigma_m^2 }\right] dm' dm \,. \\
 \end{split}
\end{eqnarray}

We generate mock subhalo detection events using a Monte Carlo method.
Firstly, we randomly sample $N$ haloes with mass in the range
$[10^{13}, 10^{14}]$ $\ms$ using the mass function of the
\textsc{eagle} reference simulation. This mass range is
consistent with the lens sample in the Sloan Lens ACS Survey
(SLAC) \citep{Vegetti2014}. For simplicity, we assume that for all the
strong lens systems, $z_l=0.3$ and $z_s=0.5$, comparable to the values
in the SLAC observations.

Using Eq.\ref{eq:RE}-\ref{eq:MR}, we calculate the velocity dispersion and 
the Einstein radius for each halo, and the corresponding mass contained within each ring,
$M_{\rm ring}$. {\color{black} We assume the dark matter haloes follow the NFW profile\citep{NFW97} with
concentration-mass relation derived by \citet{Neto2007}.    According to Eq.\ref{eq:vdisp}, the velocity 
dispersion of our lenses ranges from 160 km/s to 260 km/s, comparable to 
the lenses found in the observations \citep[e.g.][]{Sonnenfeld2013}.}
We assume that the appearance of a subhalo follows a
Poisson distribution with expectation $\mu(\alpha,\beta,m_{\rm
  c},f_E,M_{\rm ring} )$. We then sample the subhaloes according to
Eq.~\ref{eq:Pm} assuming a Gaussian measurement error with standard
deviation, $\sigma_m$, for each subhalo.

To date, the smallest subhalo mass measured using this technique is
$1.9 \pm 0.1 \times 10^8 \rm M_{\odot}$, detected with a significance
of 12$\sigma$ \citep{Vegetti2012}. In this study, we consider two
values for the minimum detection threshold, $M_{\rm low}$: $ 10^8
\ms$, the limit of current observations, and $10^7\ms$, our optimistic
expectation for future observations. {\color{black} We generate mock datasets for
both CDM and WDM with $N = 50$, 100 and 1000 host haloes with
Einstein rings.}

\section{Bayesian interference for subhalo detections}
\label{sec:bayesian}

The differences in subhalo detection rates  can be interpreted 
quantitatively using Bayesian theory. Here, we follow the formalism
developed by \citet{Vegetti2009b}, outlined below.

Assuming that subhaloes follow a Poisson distribution in a lens
system, the likelihood of finding $n_s$ subhaloes of mass, ${\bf m}$,
in an Einstein ring system can be written as:
\begin{equation}
\mathcal{L}(n_s,{\bf m} | {\bf p, q})=\frac{e^{-\mu} \mu^{n_s}}{n_s!}\prod^{n_s}_{i=1}P(m_i|{\bf p, q})\,,
\end{equation}
where the vector, ${\bf q}=\{\alpha, f_E, M_{\rm ring}, \beta,m_c\}$, gives the
parameters of the model and the vector, ${\bf p} =\{ M_{\rm min},
M_{\rm max}, M_{\rm low} \} $, contains the fixed values of the
parameters that define the minimum and maximum mass allowed by the
subhalo mass function and the threshold detection limit of a given
observation. If the errors on the measurement of subhalo mass are
Gaussian distributed with standard deviation, $\sigma_m$, $P(m_i|{\bf
  p, q})$ gives the probability of finding a subhalo with detected
mass, $m_i$, given the true subhalo mass distribution function,
$\frac{dP}{dm}\Big\vert_{\rm
  true}$.
\begin{equation}
P(m_i|{\bf p, q})=\frac{\int^{M_{\rm max}}_{M_{\rm min}} \frac{dP}{dm}\Big\vert_{\rm true} \exp\left[\frac{(m_i-m')^2}{2\sigma_m^2 }\right] dm' } 
{\int^{M_{\rm max}}_{M_{\rm low}} \int^{M_{\rm max}}_{M_{\rm min}} \frac{dP}{dm}\Big\vert_{\rm true} \exp \left[\frac{(m-m')^2}{2\sigma_m^2 }\right] dm' dm}
\end{equation}
The denominator in this equation is a normalization factor. Given $N$
Einstein ring systems, the total likelihood can be
computed as:
\begin{equation}
\mathcal{L}_{\rm tot}=\prod_{j=0}^{N}\mathcal{L}(n_{j},{\bf m_j}|\bf p,q) \,,
\end{equation}
with $n_{j}$ and ${\bf m}_j$  the number and masses of subhaloes detected in the $j$th system.

We perform a MCMC fitting to the mock lens systems.  The
  model has 5 free parameters: ${\bf q}=\{\alpha, f_E, M_{\rm
    ring}, \beta,m_c\}$. In the likelihood function, $f_E$ and $M_{\rm
    ring}$ are completely degenerate and so they cannot be determined
  separately using subhalo number counts. In a real observation, the
  strong lensing image can be used to determine the total mass within
  the thin annulus around the Einstein ring region and the stellar
  mass of the central galaxy can be obtained from multiband
  photometry. Combining these two masses fixes $M_{\rm ring}$. 
Here, we simply set $M_{\rm ring}$ to the value obtained from the
MCMC.

As mentioned earlier, the SHMF in CDM follows a power law in mass of 
exponent, $\alpha=1.9$ \citep{Springel2008, Gao2012}. We
therefore adopt a Gaussian prior for $\alpha$ with expectation 1.9 and
standard deviation 0.1. We also adopt a Gaussian prior for $\beta$
(Eq.~\ref{eq:nw_nc}) with expectation 1.3 and standard deviation 0.1.

{\color{black}  In this paper, we consider keV warm dark matter. We derive $m_c$ for a set of
WDM simulations in \citet{Lovell2014}, and find that $\log{m_c}$ increases almost linearly
with decreasing of dark matter particle mass.  We assume the probability distribution
of particle mass is uniform for keV WDM, so we adopted a flat prior for $m_c$ in log space. 
In this paper, we use the $f_E$ model in \citet{Han2015} to generate mock observations. 
In a real universe, different galaxy formation process can influence the survive of substructures. 
We thus assume conservatively for $f_E$ a uniform prior ranging from 0 to 1. We have also tried
a flat prior in log space for $f_E$ and find that the differences in posterior distribution are negligible.}

\section{Results}
\label{sec:result}

\begin{figure*}
  \includegraphics[width=\textwidth]{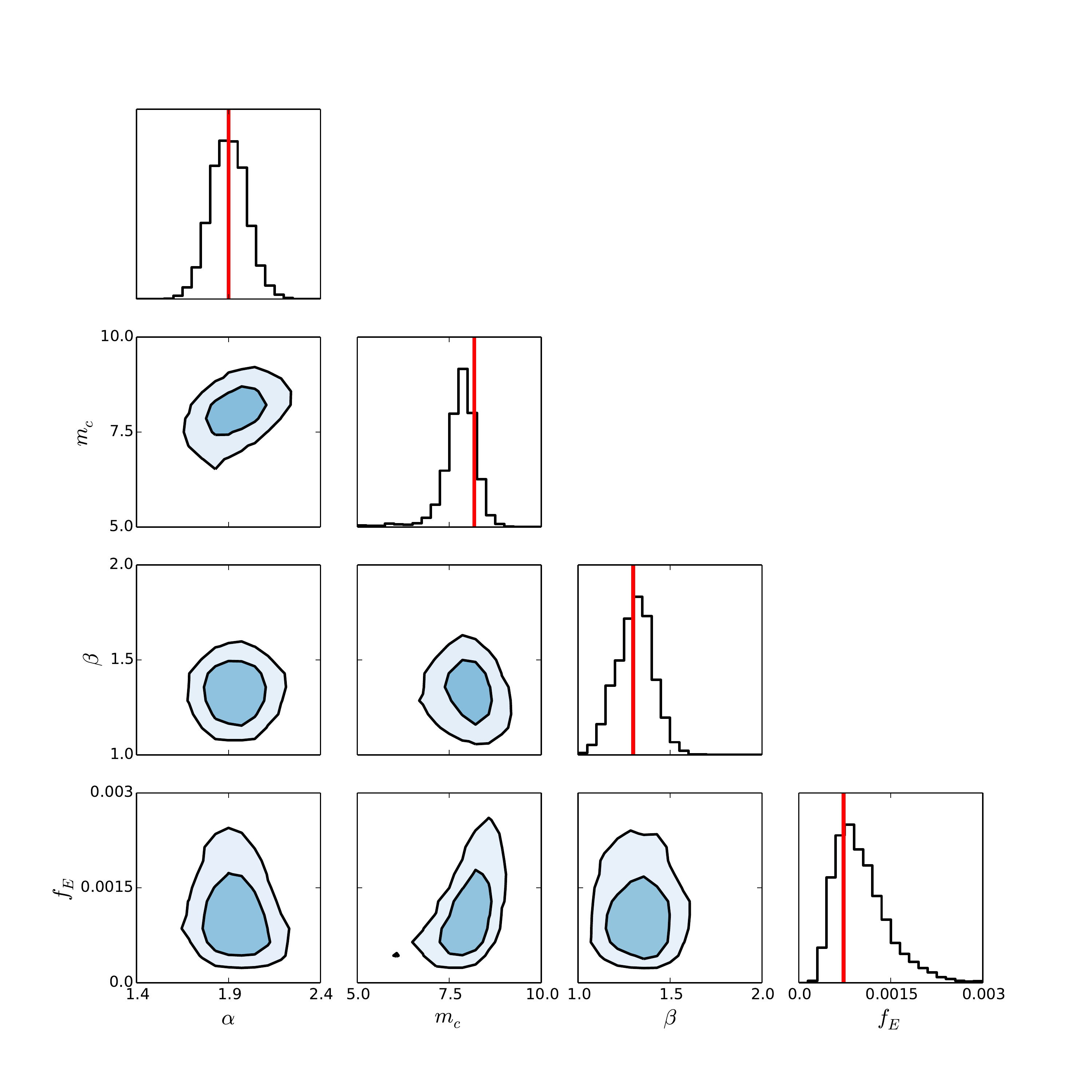}
  \caption{Parameter constraints  from 100 mock systems constructed
    using parameters appropriate to the  \textsc{coco-warm}
    simulation. The contours show the 68\%  and 95\%   confidence
    levels for the 2D posterior probability distribution of 
    the model parameters. The histograms at the end of each row show the
    marginalized 1D posterior probability distribution for each model parameter.
    The red vertical lines show the input values of each model
    parameter. The assumed detection limit is $M_{\rm low}=10^7\ms$. }
\label{fig:mcmc}
\end{figure*}

Fig.~\ref{fig:mcmc} shows the results of the MCMC analysis using 100
mock systems constructed using parameters appropriate to
\textsc{coco-warm}. Here, the input SHMF is obtained from
Eq.~\ref{eq:nw_nc} with $m_c=1.3\times10^8 \ms$. The detection limit
was set to $M_{\rm
  low}=10^7\ms$. The contours show the 68\% and 95\% confidence levels
for the 2D posterior probability distribution of model parameters,
while their marginalized 1D posterior probability distributions are
shown as histograms at the end of each row. The red vertical lines
show the input value of each parameter. The 2D contours
  indicate that the parameters, $f_E$ (the fraction of dark matter mass
  in subhaloes within the Einstein radius), and, $m_{\rm c}$ (the
  cutoff mass), are slightly
  degenerate. That is to say, the lack of small haloes in WDM can be
  partialy compensated for by a decrease in the overall amplitude of the
  SHMF. {\color{black} With a detection limit of $M_{\rm low}=10^7\ms$ and $N=100$
  systems, both $m_{\rm c}$ and $f_E$ are tightly constrained.
  Crucially, we find that with data like these one can rule out at the
  2$\sigma$ level all dark matter models with $m_c<10^{6.64}\ms$,
  which includes CDM.}

\begin{figure*}
  \includegraphics[width=\textwidth]{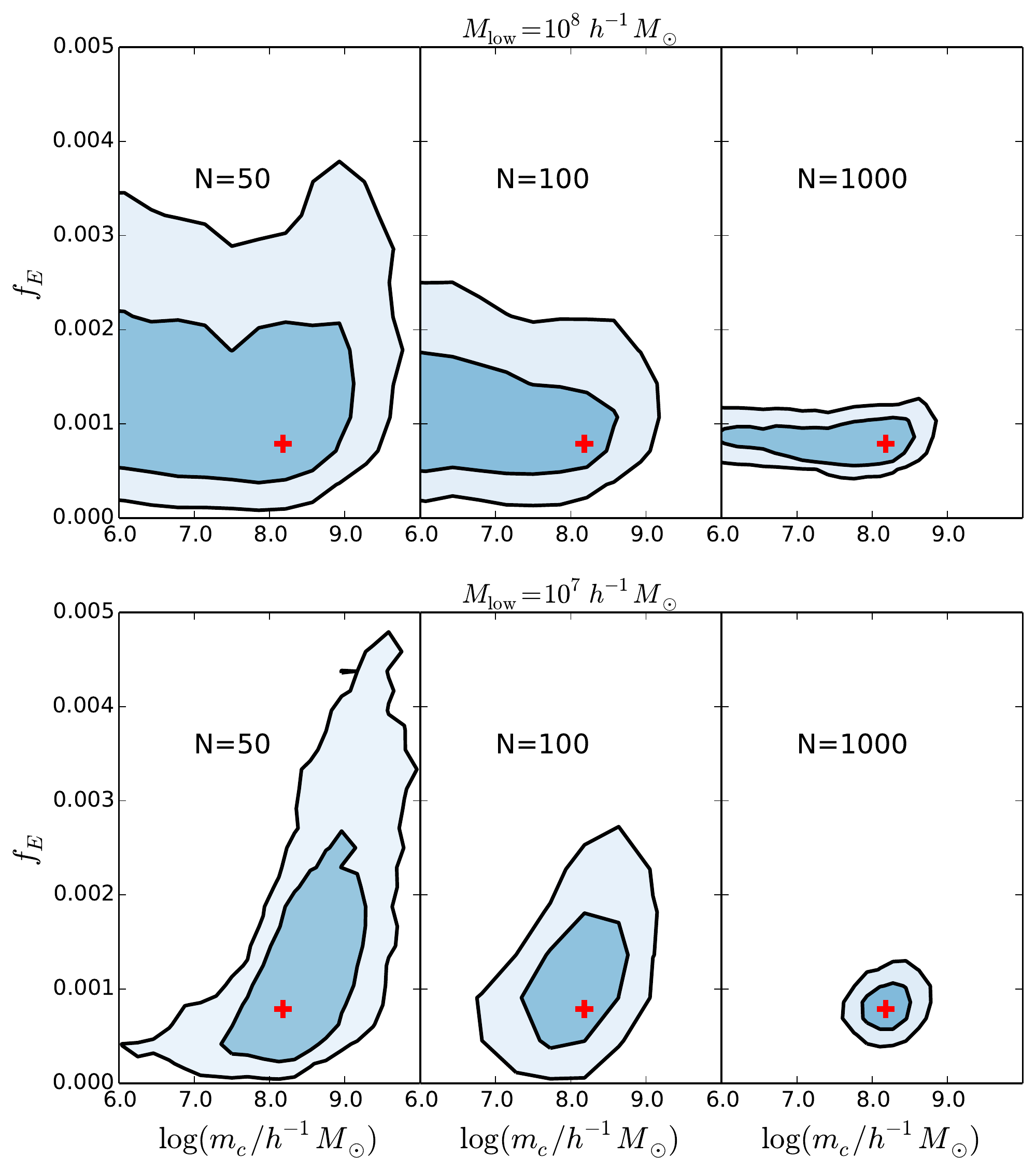}
  \caption{The constraining power on $f_E$ and $m_c$ using 50, 100 and
    1000 mock Einstein ring systems. The upper panels show  results
    for $M_{\rm low}=10^8\ms$, while the lower panels show 
    results $M_{\rm low}=10^7\ms$. The input SHMF is from
    \textsc{coco-warm}. The red crosses show the parameter values of
    \textsc{coco-warm}. }
\label{fig:mockN}
\end{figure*}

  We now explore how the number of strong lens systems, $N$, 
  affects the constraining power of the method. In
  Fig.~\ref{fig:mockN}, we show constraints on $f_E$ and $m_c$
  using 50, 100 and 1000 mock systems for detection limits of $M_{\rm
    low}=10^7\ms$ and $M_{\rm low}=10^8\ms$. The
  1$\sigma$ error on $f_E$ decreases by about a factor of 3 as $N$
  increases from 50 to 100. {\color{black} Even with $N=50$ lenses,
   one can still put constraints on the lower limit as long as subhaloes as massive as $M_{\rm
      low}=10^7\ms$ can be detected.}

\begin{figure}
\includegraphics[width=0.5\textwidth]{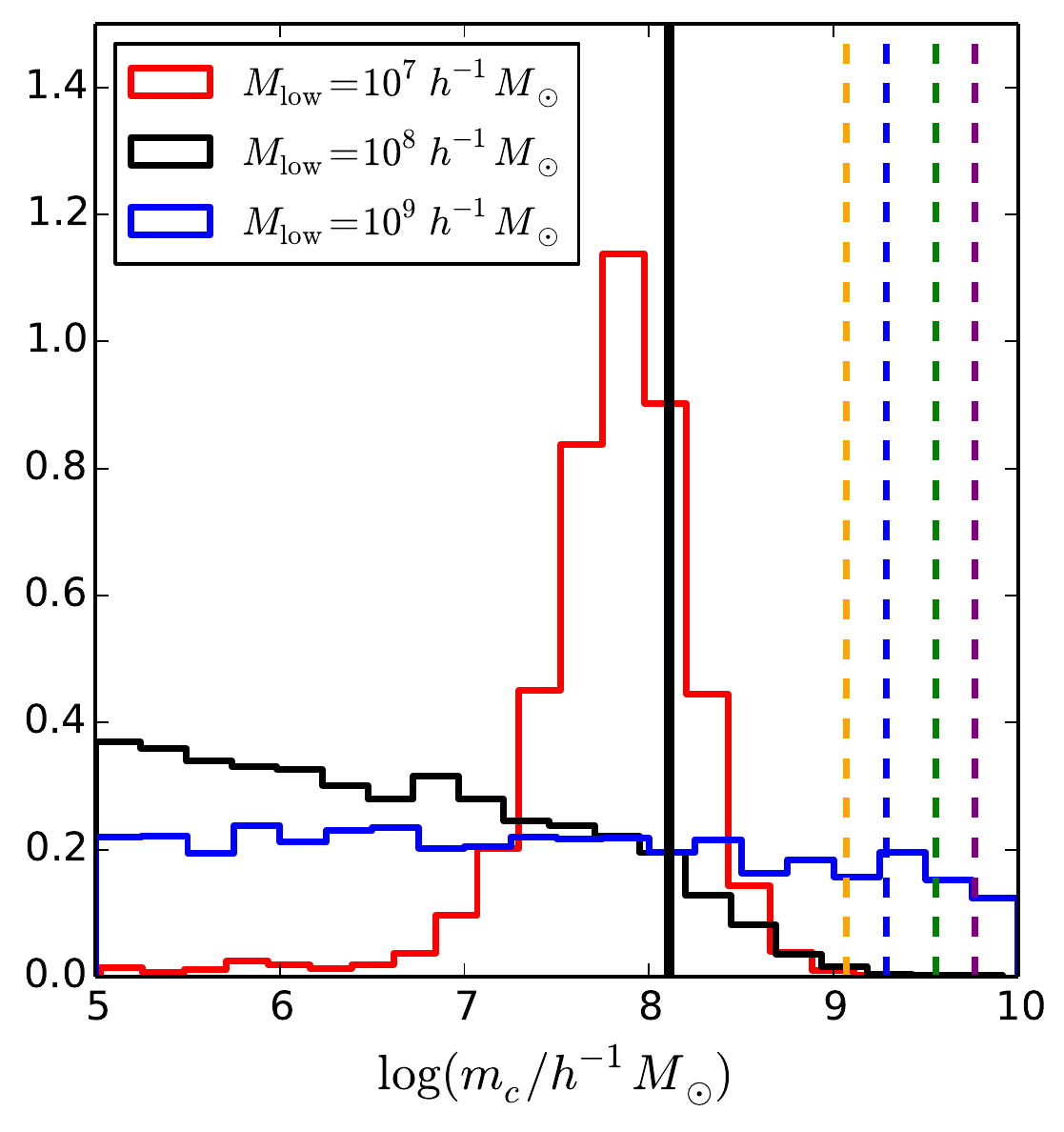}
\caption{The marginalized 1D probability distribution of $m_c$ for
  different detection mass limits with $N=100$. The mock systems are
  generated using \textsc{coco-wdm} subhalo mass function. The
  vertical black solid line shows the $m_c$ value of the 
  \textsc{coco-warm} simulation. The coloured dashed lines from left to right
  show the $m_c$ values of warm dark matter models with particle
  masses $m_{\rm WDM}=2.28,1.95,1.59,1.41$~keV respectively. }
\label{fig:WDM_ML}
\end{figure}

    The variation of the constraints on $m_c$ for different values of $M_{\rm
      low}$ is displayed in  Fig.~\ref{fig:WDM_ML}. Red, black and blue
      histograms show the marginalized 1D posterior probablity distribution of
    $m_c$, for detection limits of $M_{\rm low}=10^7\ms$, $M_{\rm
      low}=10^8\ms$ and $M_{\rm low}=10^9 \ms$ respectively.
    A detection limit of $M_{\rm low}=10^9 \ms$ hardly
    constrains the properties of the dark matter. This is not only
    because of poor detectability, but also because the number of
    subhaloes above   this mass that can be found within a host halo is intrinsically
    small. {\color{black} For $M_{\rm low}=10^8 \ms$, dark matter models with
    $m_c>10^{8.5}\ms$ are disfavoured,
    but the lower limit of $m_c$ still cannot be constrained.}
    Our results illustrate
 the vital importance of the subhalo detection threshold in distinguishing
 different dark matter models.
 
 \citet{Lovell2014} resimulated four WDM analogues of the CDM galactic
 haloes in the \textsc{Aquarius} simulations \citep{springel2009} for warmer models
 than \textsc{coco-warm}, specifically for models with power spectrum
 cutoffs corresponding to thermal relic warm particle masses of $m_{\rm
   WDM} = \left[2.28,
   1.96, 1.59, 1.41\right]$~keV. By fitting Eq.~\ref{eq:nw_nc} to
 the SHMF in each case, we can obtain values for $m_c$, which increase
 for  decreasing values of $m_{\rm WDM}$. 
 We find best-fit values of 
 $\log[m_c/(\ms)]=\left[9.07,9.28,9.55,9.76\right]$ for $m_{\rm WDM} =
 \left[2.28, 1.96, 1.59, 1.41\right]$~keV respectively. These
 values are overplotted as the dashed black lines in
 Fig.~\ref{fig:WDM_ML}. It can be seen that with $M_{\rm
   low}=10^8\ms$ one can set a strong lower limit to 
 $m_{\rm WDM}$.

\begin{figure}
  \includegraphics[width=0.5\textwidth]{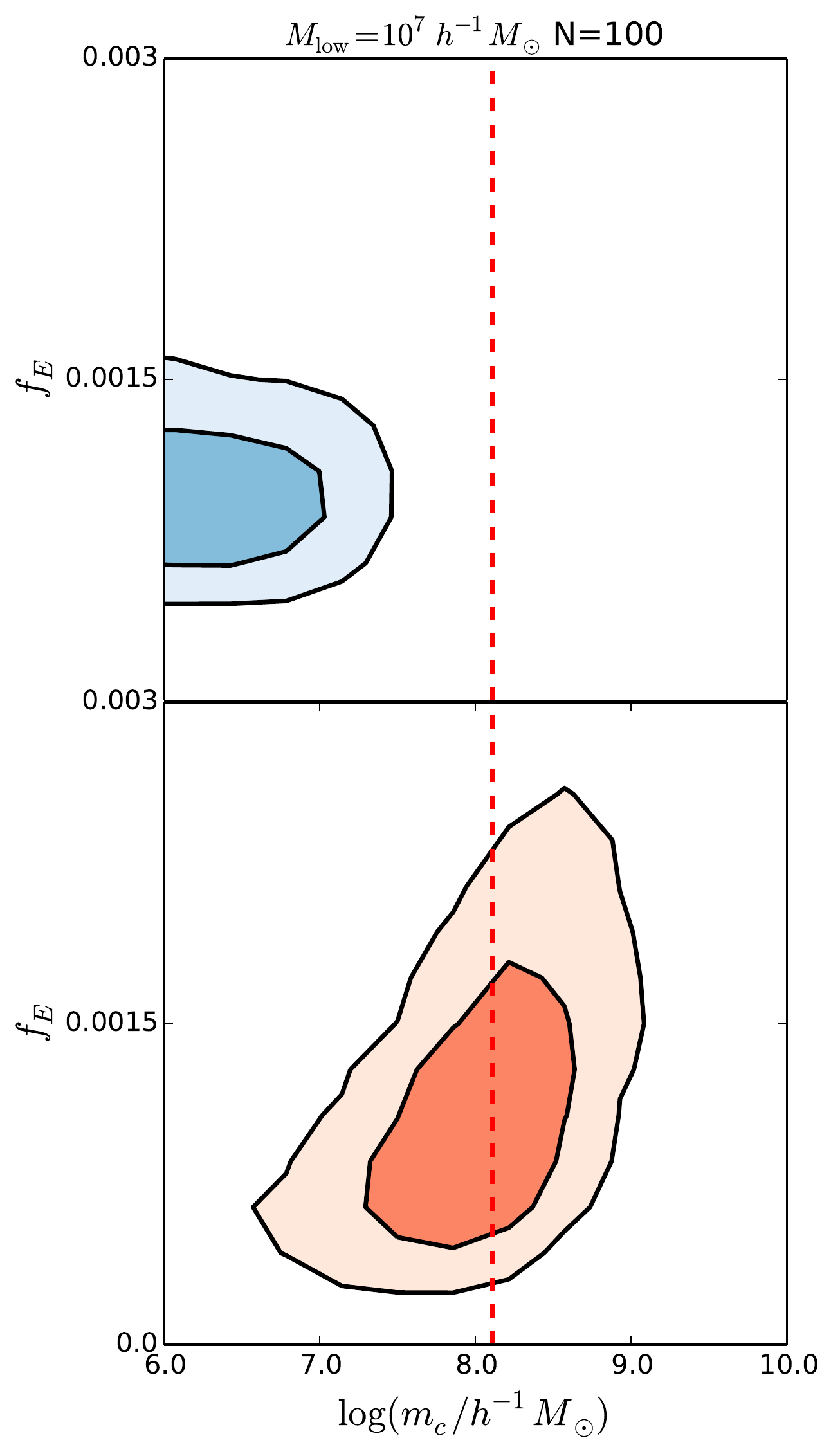}
  \caption{The 2D posterior probability distribution of $f_E$ and
    $m_c$, assuming  $N=100$, $M_{\rm low}=10^7\ms$. The lower panel 
    shows the results with input model of \textsc{coco-warm} (red), and 
    the upper panel shows the results with input mode of \textsc{coco-cold} (blue).
     The vertical dashed line shows the $m_c$ value of the \textsc{coco-warm} simulation.}
\label{fig:CDM_WDM}
\end{figure}

Finally, in Fig.~\ref{fig:CDM_WDM} we show the 2D posterior
probability distributions of $f_E$ and $m_c$ using input models of
 \textsc{coco-cold}(upper) and \textsc{coco-warm} (lower), with $N=100$ and a detection
limit of $M_{\rm low}=10^7\ms$. Encouragingly, we find that this
observational set up is sufficient to distinguish between the two
cosmologies. In other words, by observing approximately {\color{black}100 strong lens systems} with
a detection threshold of $M_{\rm low}=10^7\ms$, we could potentially
rule out the 3.3~keV thermal WDM model, which, as discussed earlier,
has a very similar power spectrum to the ``coldest'' 7~keV
sterile neutrino model. This is therefore a promising way potentially
to rule out the entire family of 7~keV sterile neutrinos as candidates
for the dark matter.

{\color{black} In table \ref{tab:mcmc}, we show the 95\% error range for recovered $m_c$ and $f_E$ from MCMC for different N and $M_{\rm low}$.}

\begin{table*}
\begin{center}
  \caption{The 95\% error range for recovered $m_c$ and $f_E$ from MCMC for different N and $M_{\rm low}$  and for CDM and WDM models.}

\begin{tabular}{l|c|c|c|c|c|c|c}
 \hline   
 &&&&\\
    
 && &WDM &&& CDM\\
 \cline{2-6}
  &&&&\\

&$\log{M_{\rm low}}=8$
&
&$\log{M_{\rm low}}=7$
&
&
&$\log{M_{\rm low}}=7$\\

   &&&&\\
&N=100
& N=1000
&N=50
& N=100
& N=1000
&N=100\\

   &&&&\\
  \hline 
   & & && \\

$f_E/0.001$ &[0.42,2.21]& [0.58,1.11]& [0.25,4.51] &[0.42,2.12] & [0.56,1.11] & [0.48,1.80]\\
 &  & && \\
 \hline
  &&&&\\
$\log{m_c}$ &$< 8.5$& $<8.5$& [ 6.8, 9.6] &[6.64,8.53] & [ 7.7, 8.5] & $< 7.6$\\
 &  & & &\\

 \hline
\end{tabular}
\label{tab:mcmc}
\end{center}
\end{table*}

\section{Summary and Discussion}
\label{sec:sum}

In this paper we have investigated the potential of strong
gravitational lensing as a diagnostic of the identity of the dark
matter. Two of the currently most plausible elementary particle
candidates for the dark matter, CDM and WDM, make very different
predictions for the number of low-mass subhaloes that survive within
larger haloes by the present day. Strong lensing is sensitive to precisely this
population since subhaloes can produce measurable distortions to Einstein
rings. 

To explore the extent to which strong lensing can constrain the
subhalo mass function, we have performed Monte-Carlo simulations to
mimic observations of haloes hosting the subhalo mass functions of the
\textsc{coco-warm} and \textsc{coco-cold} high-resolution N-body
simulations. The former has a power spectrum appropriate for a 3.3~keV
thermal relic, which happens to be a very good approximation to the
power spectrum of the {\em coldest} WDM model which is consistent with
a sterile neutrino decay interpretation of the 3.5~keV X-ray line
recently discovered in galaxies and clusters \citep{Bulbul2014a,
  Boyarsky2014a}\footnote{This model is also consistent with current
  constraints on the number of small-mass haloes at high redshift
  derived from the Lyman-$\alpha$ forest \citep{Viel2013}.}. Since the
free-streaming cutoff wavelength in the linear power spectrum of WDM
density fluctuations scales inversely with the mass of the particle,
ruling out this model by detecting subhaloes of mass below the mass
corresponding to the cutoff scale, would also rule out all other
sterile neutrino models compatible with the X-ray line.

The subhalo mass function in \textsc{coco-warm} begins to fall below
the subhalo mass function of \textsc{coco-cold} at a mass of $\sim
10^9 \ms$. The difference between the two mass functions grows to a
factor of two at $10^8 \ms$, and to an order of magnitude at
$10^7\ms$.

Our analysis, shows that both the subhalo detection limit, $M_{\rm
  low}$, and the number of observed strong lensing systems are the key
for constraining the dark matter model. Specifically, we have shown
that a sample of approximately {\color{black} 100 Einstein ring systems} with detection limit,
$M_{\rm low}=10^7\ms$, is enough clearly to distinguish between the
subhalo mass functions of \textsc{coco-warm} and
\textsc{coco-cold}. In other words, if we live in a universe in which
the dark matter predominantly consists of 7~keV sterile neutrinos,
this test would conclusively rule out CDM, whereas if we live in a
universe in which the dark matter predominantly consists of CDM, the
test would rule out all 7~keV sterile neutrino families. If the
detection limit is $10^8\ms$, the test with about 100 lenses can still
set a lower limit on the WDM particle mass, but it cannot rule out CDM.
We stress, however, that tests assuming a more
realistic sensitivity function \citep[see][]{Vegetti2014} are required
for a precise result.
 
Our results highlight the enormous potential for dark matter research
of high resolution imaging surveys to search for strong lensing
systems. Current optical surveys have found $\sim10^2$ strong lenses,
but only a fraction of them have sufficiently high quality data for a
measurement of the subhalo mass function. A few subhaloes of mass
below $10^9\ms$ have already been detected
\citep{Vegetti2010,Vegetti2012, Vegetti2014}. Currently, the lowest
subhalo mass detected in an Einstein ring, which was imaged at the
Keck telescope, is $1.9 \pm 0.1 \times 10^8 \rm M_{\odot}$
\citep{Vegetti2012}. These authors claim that the detection
sensitivity of data of this quality can reach $2\times10^7 \rm
M_{\odot}$. This is the level required to carry out the test described
in this paper.

Planned ground-based telescopes such as LSST and  space missions
 such as Euclid will increase the sample of strong
 lenses by several orders of magnitude. Euclid, for example, may be
 able to obtain high resolution images for
 $\sim 10^5$ strong lenses \citep{Pawase2014}. At the same time, the SKA survey will
 increase the sample of strong radio lenses also to $\sim 10^5$.
 Follow-up observations with VLBI may even detect $10^6 \ms$ subhaloes
 \citep{McKean2015}. Aside from direct or indirect detection of the
 dark matter particles themselves, Einstein ring systems currently
 offer the best astrophysical test of the nature of the dark matter.

 \section*{Acknowledgements}

 We thank Simona Vegetti, Qiao Wang, Richard Massey for extensive and helpful discussions of
 which significantly improved this paper. RL acknowledges NSFC grant
 (Nos.11303033,11511130054), support from the Newton Fund and Youth Innovation
 Promotion Association of CAS . CSF acknowledges the European Research
 Council Advanced Investigator grant, GA 267291, COSMIWAY.
 WAH acknowledges support from Science and Technology Facilities Council grant ST/K00090/1
and the Polish National Science Center under contract \#UMO-
2012/07/D/ST9/02785. LG acknowledges support from NSFC grants Nos.11133003 and 11425312, 
the Strategic Priority Research Program, "The Emergence of Cosmological Structure"
of the Chinese Academy of Sciences (No. XDB09000000), MPG partner Group family, and
an STFC and Newton Advanced Fellowship.
 This work was supported by the Consolidated Grant [ST/L00075X/1]
 to Durham from the Science
 and Technology Facilities Council. This work used the DiRAC Data
 Centric system at Durham University, operated by the Institute for
 Computational Cosmology on behalf of the STFC DiRAC HPC Facility
 (www.dirac.ac.uk). The DiRAC system is funded by BIS National
 E-infrastructure capital grant ST/K00042X/1, STFC capital grant
 ST/H008519/1, STFC DiRAC Operations grant ST/K003267/1, and Durham
 University. DiRAC is part of the National E-Infrastructure.

\bibliography{ref}

\end{document}